\documentclass{article}
\usepackage{url}
\usepackage{calc}
\usepackage{graphicx}
\usepackage{amssymb,amsfonts,amsmath}
\usepackage{floatflt}
\usepackage[algo2e,algoruled,vlined,longend]{algorithm2e}
\SetKw{KwOr}{or}
\SetKw{KwAnd}{and}
\SetKw{KwWhether}{Whether}
\SetKwFor{MyForEach}{for each}{do}{endfch}

\newtheorem{theorem}{Theorem}
\newtheorem{corollary}[theorem]{Corollary}
\newtheorem{proposition}[theorem]{Proposition}
\newtheorem{remark}[theorem]{Remark}

\def\N{\mbox{I\hspace{-.15em}N}}
\newcommand{\Z}{\ensuremath{\mathbb Z}}
\newcommand{\pF}[1]{\leavevmode
        \kern.1em\raise.0ex \hbox{\Z}\kern-.1em /\kern-.15em\lower.3ex
         \hbox{\ensuremath{#1}}\mbox{\Z}}
\DeclareMathOperator{\nequiv}{\mskip4mu\not\equiv\mskip4mu}

\newfont{\saffaddrit}{phvro at 8pt} % JGD 26.09.2003
\newfont{\seaddfnt}{phvr at 9pt} % JGD 26.09.2003

\newenvironment{proof}{\noindent{\bf Proof. }}{\hfill $\Box$ \smallskip}

\begin{document}
\title{Efficient polynomial time algorithms computing industrial-strength primitive roots}
\author{Jacques Dubrois\footnote{
    \mbox{Axalto},
50 Avenue Jean-Jaur\`es, B.P. 620-12
92542 Montrouge, France.
\texttt{jdubrois@axalto.com}
}
and Jean-Guillaume Dumas 
\footnote{
    \mbox{Universit\'e Joseph Fourier,}
    \mbox{Laboratoire  J. Kuntzmann, umr CNRS 5224,}
    51 av. des Math\'ematiques.
    \mbox{B.P. 53 X, F38041 Grenoble, France.}
\texttt{Jean-Guillaume.Dumas@imag.fr}
}}
%\ead[url]{www-lmc.imag.fr/lmc-mosaic/Jean-Guillaume.Dumas@imag.fr}
\maketitle
 \begin{abstract}
 E. Bach, following an idea of T. Itoh, 
 has shown how to build a small set of numbers modulo a prime p such
 that at least one element of this set is a generator of
 $\pF{p}$. 
 E. Bach suggests also that at least
 half of his set should be generators.
 We show here that a slight variant of this set can indeed be made to
 contain a ratio of primitive roots as close to 1 as necessary.
 %We thus derive several algorithms computing primitive roots correct with very high probability in polynomial time. 
 In particular we present an asymptotically $O^{\sim}\left( \sqrt{\frac{1}{\epsilon}}log^{1.5}(p) + \log^2(p)\right)$
 algorithm providing primitive roots of $p$ with probability of correctness greater
 than $1-\epsilon$ and several $O(log^\alpha(p))$, $\alpha \leq 5.23$, algorithms computing "Industrial-strength" primitive roots.
 % with probabilities e.g. 
 % greater than the probability of "hardware malfunctions".
 \end{abstract}

% \begin{keyword}
% computational complexity
% \sep cryptography \sep randomized algorithms 
% %  \sep algorithms \sep analysis of algorithms
% % % keywords here, in the form: keyword \sep keyword
% % % PACS codes here, in the form: \PACS code \sep code
% % \PACS 
% \end{keyword}
% \end{frontmatter}
%
\section{Introduction}
Primitive roots are generators of the multiplicative group of the
invertibles of a finite field. We focus in this paper only on prime
finite fields, but the proposed algorithms can work over
extension fields or other multiplicative groups.

Primitive roots are of intrinsic use e.g. for secret key exchange
(Diffie-Hellman), pseudo random generators (Blum-Micali) or primality certification.
%% Mersenne (Project 8, florida)
The classical method of generation of
such generators is by trial, test and error. Indeed within a prime field
with $p$ elements they are quite numerous ($\phi(\phi(p))=\phi(p-1)$
among $p-1$ invertibles are
generators. 

The problem resides in the test to decide whether a number $g$ is a
generator or not. The first idea is to test every $g^i$ for $i=1..p-1$
looking for matches. Unfortunately this is exponential in the size of
$p$. An acceleration is then to factor $p-1$ and test
whether one of the $g^{\frac{p-1}{q}}$ is $1$ for $q$ a divisor of
$p-1$. If this is the case then $g$ is obviously not a generator. On
the contrary, one has proved that the only possible order of $g$ is
$p-1$. Unfortunately again, factorization is still not a polynomial
time process: no polynomial time algorithm computing
primitive roots is known.

However, there exists polynomial time methods isolating a polynomial
size set of numbers containing at least one primitive root. 
Shoup's \cite{Shoup:1992:SPR} algorithm is such a
method. 
Elliot and Murata \cite{Elliott:1997:ALPR} also gave
polynomial 
lower bounds on the least primitive root modulo p. One can also
generate elements with exponentially large order even though not being
primitive roots \cite{VonzurGathen:1998:OGP}. Our method is in between
those two approaches.

As reported by Bach \cite{Bach:1997:sppr}, Itoh's breakthrough was
to use only a partial factorization of $p-1$ to produce primitive
roots with high probability \cite{Itoh:2001:PPR}.
Bach then used this idea of partial
factorization to give the actually smallest known set, deterministically
containing one primitive root\cite{Bach:1997:sppr}, if the Extended
Riemann Hypothesis is true. Moreover, he
suggested that his set contained at least half primitive roots.

In this paper, we propose to use a combination of Itoh's and Bach's
algorithms producing a polynomial time algorithm generating primitive
roots with a very small probability of failure (without the ERH). 
Such generated numbers
will be denoted by ``Industrial-strength'' primitive roots.
We also have a guaranteed lower bound on the order of
the produced elements.
In this paper, we analyze the actual ratio of
primitive roots within a variant of Bach's full set. 
As this ratio is close to
$1$, both in theory and even more in practice, 
selecting a random element within this set produces a fast and
effective method computing primitive roots. 
%The probability of
%failure, i.e. of selecting a non primitive root can therefore be made
%arbitrarily low. \\

We present in section \ref{sec:algo} our algorithm and the main
theorem counting this ratio. 
Then practical implementation details and effective ratios
are discussed section \ref{sec:heuristic}.
 We conclude section
 \ref{sec:applis}
 with applications of primitive root generation, accelerated by our
 probabilistic method. Among this applications are Diffie-Hellman key exchange, 
 ElGamal cryptosystem, Blum-Micali pseudo random bit generation, 
  and a new
  probabilistic primality test based on Lucas' deterministic procedure. This test
 uses both the analysis of the first sections and the composite case.
%
%\newpage
\section{The variant of Itoh/Bach's algorithm}\label{sec:algo}
The salient features
of our approach when compared to Bach's are that:
\begin{enumerate}\vspace{-2ex}
\item We partially factor, but {\em with known lower bound on the
    remaining factors}.
\item {\em We do not require the primality} of the chosen elements.
\item {\em Random elements are drawn from
the whole set of candidates instead of only
  from the first ones}.
\end{enumerate}\vspace{-2ex}
Now, when compared to Itoh's method, we use a deterministic process
producing a number with a very high order and 
which has a high probability of being primitive.
On the contrary, Itoh selects a random element but uses a polynomial
process to prove that this number is a primitive root with high
probability \cite{Itoh:2001:PPR}. 
\begin{algorithm2e}[htb]\caption{Probabilistic Primitive Root}
\raggedright
\label{algo1}
  \KwIn{A prime $p\geq 3$ and a failure probability $0 < \epsilon < 1$.}
  \KwOut{A number, primitive root with probability greater than $1-\varepsilon$.}
  \Begin{
    Compute $B$ such that
    $(1+\frac{2}{p-1})(1-\frac{1}{B})^{\log_B{\frac{p-1}{2}}} = 1-\varepsilon$.\\
    Partially factor $p-1=2^{e_1}p_2^{e_2}. . . . .p_h^{e_h}Q$  
    ({\em $p_i<B$ and $Q$ has no factor $<B$}).\\
    \MyForEach{$1\leq i \leq h$}{
      By trial and error, randomly choose $\alpha_i$ verifying: 
      $\alpha_{i}^{\frac{p-1}{p_i}}\nequiv 1\;(mod\;p)$.\\
    }
    Set $a \equiv \prod\limits_{i=1}^h \alpha_{i}^{\frac{p-1}{p_i^{e_i}}}\;(mod\;p)$.\\
    \eIf{Factorization is complete}{
      Set Probability of correctness to $1$ and
      \KwRet{$a$}.
    }{
      Refine Probability of correctness to
      $(1+\frac{1}{Q-1})(1-\frac{1}{B})^{\log_B{Q}}$.\\
      Randomly choose $b$ verifying:
      $b^{\frac{p-1}{Q}} \nequiv 1$ and
      \KwRet{$g \equiv  ab^{\frac{p-1}{Q}} (mod\;p)$.}
    }
  }
\end{algorithm2e}
The difference here is that 
we use low order terms to build higher order elements whereas
Itoh discards the randomly chosen candidates and restarts all over at
each failure.
%
%As we show next, with this requirements we are able to prove very high
%probabilities to find primitive roots.
%We now prove that this algorithm is correct and give its running complexity.
Therefore we first compute the ratio of primitive roots
within the set. We
have found afterwards that Itoh, independently and differently, proves 
quite the same within his \cite[Theorem 1]{Itoh:2001:PPR}.
\begin{theorem}\label{th:prop}
At least
$\frac{\phi(Q)}{Q-1}$ of the returned values of
Algorithm \ref{algo1} are primitive roots.
\end{theorem}
\begin{proof}
We let $p-1=kQ$. In algorithm \ref{algo1}, the order of $a$ is $(p-1)/Q=k$
(see \cite{Bach:1997:sppr}).
We partition ${Z/pZ}^*$ by $S$ and $T$ where
$$S = \{b\in {Z/pZ}^*: b^k \nequiv 1
  (mod\;p)\}
~~\text{and}~~
T = \{b\in {Z/pZ}^* :  b^k \equiv 1
  (mod\;p)\}
$$
and let $U= \{b\in {Z/pZ}^*: b^k ~\text{has order}~ Q\}$.
Note that for any $x \in {Z/pZ}^*$ of order $n$ and any 
$y\in {Z/pZ}^*$ of order $m$, if $gcd(n,m)=1$ then the order of 
$z \equiv xy (mod\;p)$ is $nm$.
Thus for any $b \in U$ it follows that $g \equiv ab^k (mod\;p)$ has
order $p-1$.
Since $U \subseteq S$, we have that $\frac{|U|}{|S|}$ of the returned
values of algorithm \ref{algo1} are primitive roots.

We thus now count the number of elements of $U$ and $S$.
On the one hand, we fix arbitrarily a primitive root $\tilde{g} \in {Z/pZ}^*$ and
define $E = \{i: 0 \leq i \leq Q ~\text{and}~
gcd(i,Q)=1\}$. $|E|=\varphi(Q)$ and it is not difficult to see that 
$U = \{\tilde{g}^{i+jQ} : i\in E~\text{and}~0\leq j \leq k-1\}$.
This implies that $|U|=k\varphi(Q)$. 

On the other hand, we have
$T=\{\tilde{g}^{0},\tilde{g}^{Q},\ldots,\tilde{g}^{(k-1)Q}\}$. The
  partitioning therefore gives $|S|=|{Z/pZ}^*|-|T|=p-1-k$. 
We thus conclude that 
$\frac{|U|}{|S|} = \frac{ k \phi(Q)}{p-1-k}=\frac{\phi(Q)}{Q-1}$.
\end{proof}
%
%\bigskip
%
\begin{corollary}\label{dem:prob}
Algorithm \ref{algo1} is correct
and, when Pollard's rho algorithm is used, has an average running time of
$O\left(\sqrt{\frac{1}{\varepsilon}}\log^{2.5}(p)+\log^3(p)\log(\log(p))\right)
\footnote{Using fast integer arithmetic 
%\cite[Corollary 11.10]{VonzurGathen:1999:MCA} 
this can become~:\\
$O\left( \sqrt{\frac{1}{\varepsilon}}
\log^{1.5}(p)\log^2(\log(p))\log(\log(\log(p)))
+ \right.$
$ \left. \log^2(p)\log^2(\log(p))\log(\log(\log(p))) \right)
$ ;  but the worst case complexity is 
$O\left(\frac{1}{\varepsilon}\log^{3}(p)+\log^4(p)\log(\log(p))\right)$.}.
$
\end{corollary}
\begin{proof} We first need to show that $\frac{\phi(Q)}{Q-1}>1-\varepsilon$.
Let $Q=\prod\limits_{i=1}^{\omega(Q)} {q_i}^{f_i}$ where
${\omega(Q)}$ is the number of distinct prime factors of $Q$.
Then $\phi(Q)=\prod\limits_{i=1}^{\omega(Q)} \phi({q_i}^{f_i})=Q\prod\limits_{i=1}^{\omega(Q)} (1-\frac{1}{q_i})$.
Thus $\frac{\phi(Q)}{Q-1}= (1+\frac{1}{Q-1})\prod\limits_{i=1}^{\omega(Q)} (1-\frac{1}{q_i})$.
Now, since any factor of $Q$ is bigger than $B$, we have:
$\prod\limits_{i=1}^{\omega(Q)}
(1-\frac{1}{q_i})>\prod\limits_{i=1}^{\omega(Q)}
(1-\frac{1}{B})=(1-\frac{1}{B})^{\omega(Q)}.$
To conclude, we minor $\omega(Q)$ by $\log_B(Q)$. This gives the probability
refinement\footnote{%
Note that one can dynamically refine $B$ as more 
factors of $p-1$ are known.}. 
Since $Q$ is not known at the beginning, one can minor it
there by $\frac{p-1}{2}$ since $p-1$ must be even whenever $p\geq3$.
Now for the complexity.
For the computation of $B$, we use a Newton-Raphson's
approximation.
The second step depends on the factorization method. Both complexities
here are given by the application of Pollard's rho algorithm.
Indeed Pollard's rho would require at worst
$L = 2\lceil B \rceil$ loops and $L = O(\sqrt{B})$ on the average
thanks to the birthday paradox. 
Now each loop of Pollard's rho is a squaring and a gcd, both of
complexity $O(\log^2{p})$.

Then we need to bound $B$ with respect to $\varepsilon$.
We let $h=(p-1)/2$ and $B^{*}=min\{ln(h)/\varepsilon;h\}$ and consider
$f_h(\varepsilon)=(1-1/B^{*})^{\log_{B^{*}}(h)}-(1-\varepsilon)$. Then
{\small $$f_h(\varepsilon) = \left(1-\frac{1}{ln(B^*)}\right)\varepsilon + 
\frac{1}{2ln(B^*)}\left(\frac{1}{ln(B^*)}-\frac{1}{ln(h)}\right)\varepsilon^2
%+ \frac{1}{6ln(B^*)}\left(-\frac{1}{ln(B^*)^2}-\frac{2}{ln(h)^2}+\frac{3}{ln(B^*)ln(h)}\right)\varepsilon^3
%+ o(\varepsilon^4)
+ O\left(\frac{\varepsilon^3}{6ln(B^*)^3}\right)
$$}
is strictly positive as soon as $B^*\geq 3$. 
This proves that $1-\varepsilon < (1-1/B^{*})^{\log_{B^{*}}(h)}$. Now,
since $(1-1/B)^{\log_B(h)}$ is decreasing in $B$, this shows that 
$B$ such that
$(1+\frac{2}{p-1})(1-\frac{1}{B})^{\log_B{\frac{p-1}{2}}} =
1-\varepsilon$ satisfies $B < B^* \leq \frac{ ln(h)}{\varepsilon}$.

For the remaining steps, there is at worst $\log{p}$ distinct factors,
thus $\log{p}$ distinct $\alpha_i$, but only $\log{\log{p}}$ on the average
\cite[Theorem 430]{Hardy:1979:ITN}.
Each one requires a modular exponentiation
which can be performed with $O(\log^3{p})$ operations using recursive squaring.
Now, to get a correct $\alpha_i$, at most $O(\log\log{p})$
trials should be necessary
(see e.g. \cite[Theorem 6.18]{Wagstaff:2003:CNTC}).
However, by an argument similar to that of theorem \ref{th:prop},
less than $1 - \frac{1}{p_i}$ of the $\alpha_i$ are such
that $\alpha_i^{\frac{p-1}{p_i}} \equiv 1$. This gives an average number of
trials of $1 + \frac{1}{p_i}$, which is bounded by a constant. This
gives $\log \times \log^3 \times \log\log$ in the worst case (distinct
factors $\times$ exponentiation $\times$ number of trials) and only
$\log\log \times \log^3 \times 2$ on the average.
\end{proof}%\vspace{-3em}

%%%%%%%%%%%%%%%%%%%%%%%%%%%%%%%%%%%%%%%%%%%%%%%%%%%%%%%%%%%%
%%% Local Variables:
%%% mode: latex
%%% TeX-master: "polypr"
%%% End:
 %% Algorithm and theorems
%
%\newpage
%
\section{About the number of prime divisors}\label{sec:omega}
In the previous section, we have seen that the probability to get a
primitive root out of our algorithm is greater than 
$\left(1-\frac{1}{B}\right)^{\omega(Q)}$ for $Q$ the remaining
unfactored part with no divisors less than $B$. 
The running time of the algorithm, and in particular its
non-polynomial behavior depends on $B$ and on $\omega$. In practice,
$\omega$ is quite small in general. The problem is that
the bound we used in
the preceding section, $\log_B(p-1)$, is then much too large.  
In this section, we thus provide tighter probability estimates
for some small $B$ and large $Q$.
%
%% B:=2^10  ==>  f=1.0956447776819900726
%% B:=2^15  ==>  f=
%% B:=2^20  ==>  f=1.0561363740519441931
\begin{theorem} 
Let $B \in \N$, $Q \in \N$ such that no prime lower than $B$ divides
$Q$ then:
\begin{align}
  \omega(Q) \leq ~ & \log_B(Q) & \forall B \geq ~&2\label{eq:logB}\\
\omega(Q) \leq ~ & \frac{1.0956448}{\log_B( ln(Q) )} \log_B(Q) & \forall B \geq ~& 2^{10}\label{eq:f10}\\
\omega(Q) \leq ~ & \frac{1.0808280}{\log_B( ln(Q) )} \log_B(Q) & \forall B \geq ~& 2^{15}\label{eq:f15}\\
\omega(Q) \leq ~ & \frac{1.0561364}{\log_B( ln(Q) )} \log_B(Q) & \forall B \geq ~& 2^{20}\label{eq:f20}
\end{align}
\end{theorem}
\begin{proof}
Of course, (\ref{eq:logB}) is a large upper bound on the number of
divisors of $Q$ and therefore a bound on the number of prime
divisors. Now for the other bounds, we refine Robin's bound on $\omega$
\cite[Theorem 11]{Robin:1983:omega}: which is $\omega(n)\leq \frac{1.3841}{ln(ln(n))} ln(n)$.
Let $N_k = \prod_{i=1}^k p_i$ where $p_i$ is
the i-th prime. Now, we let $k$ be such that
$\frac{N_k}{N_{\pi(B)}} \leq Q < \frac{N_{k+1}}{N_{\pi(B)}}$. Then 
$\omega(Q) \leq \omega\left(\frac{N_k}{N_{\pi(B)}} \right) = k -
\pi(B)$ 
since no prime less than $B$ can divide
$Q$. 
We then combine this with the fact that $X \hookrightarrow \frac{ln(X)}{X}$ is
decreasing for $(X>e)$, to get: 
$\omega(Q) \leq \frac{F(k,B)}{\log_B( ln(Q) )} \log_B(Q)$ where
{\small $F(k,B) = \frac{(k -\pi(B))\log\left(\log\left(\frac{N_k}{N_{\pi(B)}} \right)\right)}{\log\left(\frac{N_k}{N_{\pi(B)}} \right)} $}.
We then replace both $N_k$ in $F(k,B)$ using e.g. classical bounds on
$\theta(p_k) = ln(N_k)$ \cite[Theorems 7 \& 8]{Robin:1983:omega}:
\begin{align}
\theta(p_k) \geq & k\left(ln~k+ln~ln~k-1+\frac{ln~ln~k-2.1454}{ln~k}\right)
\\
\theta(p_k) \leq & k\left(ln~k+ln~ln~k-1+\frac{ln~ln~k-1.9185}{ln~k}\right)
\end{align}
We therefore obtain a function $\widetilde{F}(k,B)$ explicit in $k$ and $B$.
The values given in the theorem are the numerically
computed maximal values
of $\widetilde{F}(k,B)$ as a function of $k$ for 
$B \in \{ 2^{10}, 2^{15}, 2^{20} \}$. The claim then follows from the fact that
$\widetilde{F}(k,B)$ is decreasing in $B$.
\end{proof}
It is noticeable that the last estimates are more interesting than
$\log_B(Q)$ only when $B^{\widetilde{F}(k,B)} < ln(Q)$. Those
estimates are then only useful for very 
large $Q$ (e.g. more than $10^5$ bits for $B=2^{15}$).

%%%%%%%%%%%%%%%%%%%%%%%%%%%%%%%%%%%%%%%%%%%%%%%%%%%%%%%%%%%%
%%% Local Variables:
%%% mode: latex
%%% TeX-master: "polypr"
%%% End:
 
  %% number of prime divisors
%
%\newpage
%
\section{Industrial-strength primitive roots}\label{sec:heuristic}
\begin{figure}[htbp]
\includegraphics[width=\textwidth]{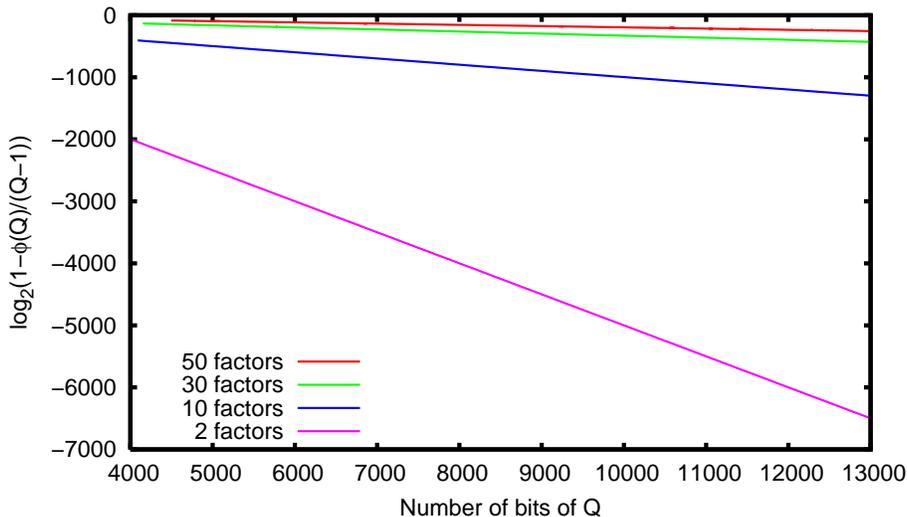}
\caption{Actual probability of failure of Algorithm \ref{algo1} with $L=2^{20}$}\label{fig:factors}
\end{figure}

Of course, the only problem with this algorithm is that it is not
polynomial. 
Indeed the partial factorization up to factors of any
given size is still exponential. This gives the non polynomial
factor $\sqrt{\frac{1}{\varepsilon}}$. Other factoring algorithms
with better complexity could also be used, provided they can
guarantee a bound on the unfound factors.
For that reason, we propose another algorithm with an attainable
number of loops for the partial factorization.
Therefore, the algorithm is efficient and we provide experimental data
showing that it also has a very good behavior with respect to the
probabilities:
\begin{center}
{\em Heuristic 2: Apply Algorithm \ref{algo1} with $B \leq
  \log^2(p)\log^2(\log(p))$.}
\end{center}
With Pollard's rho factoring, the algorithm 
   has  now an average bit polynomial complexity of : $O\left(\log^3(p)\log(\log(p))\right)$
(just replace $B$ by $\log^2(p)\log^2(\log(p))$ and use $L=\sqrt{B}$).
In practice, {\em $L$~could be chosen not higher than  a million}:
in figures
\ref{fig:factors} we choose $Q$ with known factorization and compute
$\frac{\phi(Q)}{Q-1}$ ; 
\begin{figure}[htbp]
\includegraphics[width=\textwidth]{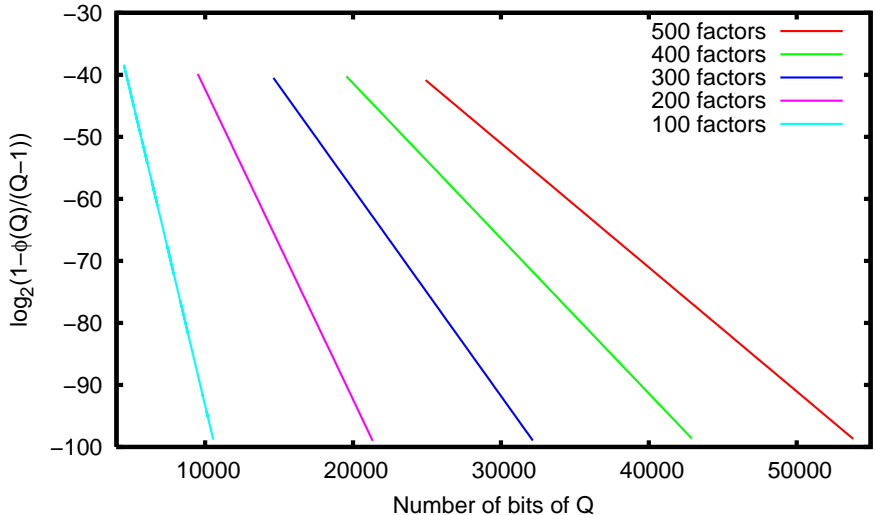}
\caption{Actual probability of failure for $Q$ with many distinct factors}\label{fig:fzoom}
%\caption{Actual probability of failure (powers of 2) with
%  $L=2^{20}$}\label{fig:factors}
\end{figure}
the experimental data then shows that in practice no probability less
than $1-2^{-40}$ is possible even with $L$ as small as $2^{20}$.\\
Provided that one is ready to accept a fixed probability, 
further improvements on the asymptotic complexity can be
made. Indeed, D. Knuth said {\em "For the probability less than
  $(\frac{1}{4})^{25}$ that such a 25-times-in-row procedures gives
  the wrong information about n. 
%   This is less than one chance in a
%   quadrillion; even if we certified a billion different primes with
%   such a procedure, the expected number of mistakes would be less than
%   1/1000000. 
  It's much more likely that our computer has dropped a bit
  in its calculations, due to hardware malfunctions or cosmic
  radiations, than that algorithm P has repeatedly guessed
  wrong."}\footnote{
More precisely, cosmic rays only can be responsible for 
$10^5$ software errors in $10^9$ chip-hours at sea
level\cite{OGorman:1996:FTC} .
At 1GHz, this makes 1 error every $2^{55}$ computations.}
% \cite{Knuth:1997:SA}. 
We thus provide a version of our algorithm guaranteeing
that the probability of incorrect answer is lower than $2^{-40}$:
\begin{center}
{\em Algorithm 3: If $p$ is small ($p < 45 171 967$), factor $p-1$
completely, otherwise apply Algorithm \ref{algo1} with 
$B =log^{5.298514}{p}$}.
\end{center}
With Pollard's rho factoring, the average asymptotic bit complexity is
then $O(\log^{4.649257}{p})$:
%\footnote{The worst
%  case exponents is $8.459536$.
Factoring numbers lower
than $45 171 967$, takes constant time.
Now for larger primes and $B=\log^{\alpha}(p)$, 
we just remark that
$(1+\frac{2}{p-1})(1-\frac{1}{B})^{\log_B{\frac{p-1}{2}}}$ is
increasing in $p$, 
so that it is bounded by its first value. 
Numerical approximation of $\alpha$ so that the
latter is $1-2^{-40}$ gives $5.298514$. The complexity
exponent follows as it is $2+\frac{\alpha}{2}$.
One can also apply the same arguments e.g. for a probability \mbox{$1-2^{-55}$}
and factoring all primes $p<2^{512}$ (since $513$-bit numbers are nowadays factorizable), then slightly degrading the complexity to $O(log^{5.229921}{p})$.
We have thus proved that a probability of at least $1-2^{-40}$
can always be guaranteed. In other words,
our algorithm is
able to efficiently produce ``industrial-strength'' primitive
roots.
%
%
% \newlength{\moitie}
% \setlength{\moitie}{\columnwidth*5/9}
% \begin{floatingfigure}[htb]{\moitie}\vspace{-1.6em}
%\includegraphics[width=\moitie,height=\columnwidth*4/9]{Pictures/comparaison-time.eps}
\begin{figure}[htb]
\includegraphics[width=\textwidth]{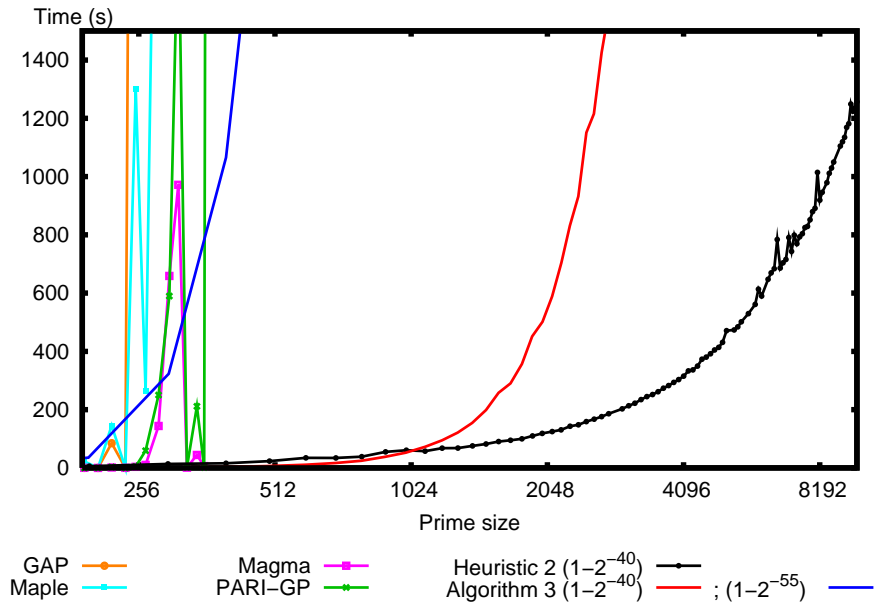}
\caption{Generations of primitive
  roots}\label{fig:pari}
%\end{floatingfigure}
\end{figure}
This is for instance illustrated when
comparing our algorithm, implemented in C++ with GMP,
to existing software
(Maple 9.5, Pari-GP, GAP 4r4 and Magma
2.11)\footnote{\url{swox.com/gmp}, \url{maplesoft.com}, \url{pari.math.u-bordeaux.fr}, \url{gap-system.org}, \url{magma.maths.usyd.edu.au}}
on an Intel PIV 2.4GHz. 
This comparison is shown on figure \ref{fig:pari}.
Of course, the comparison is not fair as other softwares are always
factoring $p-1$ completely. Still we can see the progress in
primitive root generation that our algorithm has enabled.
%%%%%%%%%%%%%%%%%%%%%%%%%%%%%%%%%%%%%%%%%%%%%%%%%%%%%%%%%%%%
%%% Local Variables:
%%% mode: latex
%%% TeX-master: "polypr"
%%% End:
 %% Efficient Polynomial time
%
%\newpage
\section{Analysis of the algorithm for composite numbers}\label{sec:compos}
In this section we propose an analysis of the behavior of the
algorithm for composite numbers. 
Indeed, our algorithm can also be used to produce high, if not
maximal, order element modulo a composite number.
This analysis is also used section
\ref{ssec:lucas} for the probabilistic primality test.
It is well known that there exists primitive roots for every number of
the form $2$, $4$, $p^k$ or $2p^k$ with $p$ an odd prime. On the other
hand, Euler's theorem states that every invertible $a \in \pF{p}^*$
satisfies $a^{\varphi(n)} \equiv 1 [n]$. Thus, for composite
numbers $n$ not possessing primitive roots, $\varphi(n)$ is not a
possible order of an invertible.
We therefore use $\lambda(m)$,
Carmichael's lambda function, the maximal order of an invertible
element in the multiplicative group ($\pF{p}^*$, $\times$).
See e.g. \cite{Knuth:1997:TAoCPSA,Erdos:1991:CLF,Bach:1996:ANTEA}, for more details.
%We therefore use 
%Carmichael's lambda function, the maximal order of an invertible, 
%defined e.g. in \cite{Knuth:1997:SA,Erdos:1991:CF,Bach:1996:ANTEA}:
%
%\begin{definition}
%$\lambda(m)$ is the maximal order of an invertible element in the
%multiplicative group ($\pF{p}^*$, $\times$).
%\end{definition}
%
Of course, $\lambda$ and $\varphi$ coincide for $2$, $4$, $p^k$ and
$2p^k$, for $p$ and odd prime. Then $\lambda(2^e)=2^{e-2}$ for $e\geq
3$.  Now, for the other cases, since 
$\varphi \left( \prod p_i^{k_i} \right) = \prod (p_i-1)p_i^{k_i-1}$
for distinct primes $p_i$, we obtain this similar formula for
$\lambda$: 
$\lambda \left( \prod p_i^{k_i} \right) = lcm \lbrace \lambda (p_i^{k_i}) \rbrace$.
Eventually, we also obtain this corollary of Euler's theorem:
\begin{corollary}
Every invertible $a$ within
$\pF{p}^*$ satisfies $a^{\lambda(n)} \equiv 1 [n]$.
\end{corollary}
\begin{proof}
$n=\prod p_i^{e_i}$ for distinct primes $p_i$.
Then $\varphi(p_i^{e_i})$ divides $\lambda(n)$. This, together with
Euler's theorem shows that  $a^{\lambda(n)} \equiv 1 [p_i^{e_i}]$. 
The Chinese theorem thus implies that the latter is also true modulo the
product of the $p_i^{e_i}$.
\end{proof}
This corollary shows that the order of any invertible must divide
$\lambda(n)$. For $n$ prime, the number of invertibles having
order $d | n-1$ is exactly $\varphi(d)$ so that 
$\sum_{d|k} \varphi(d) = k$ for $k | n-1$. We have the following
analogue for $n$ a composite number:
\begin{proposition}
The number of invertibles having
order $d | \lambda(n)$ is 
$\sum_{S_d} \prod_{j=1}^{\omega} \varphi(d_j)$
for $n= p_1^{e_1} \ldots p_{\omega}^{e_{\omega}}$
and $S_d = \lbrace (d_1,\ldots,d_\omega)$ s.t.
$d_j|\varphi(p_j^{e_j})$ and $lcm \lbrace d_j \rbrace = d \rbrace$.
\end{proposition}
\begin{proof}By the Chinese theorem, an element has order $d$ if and
  only if the lcm of its orders modulo the $p_j^{e_j}$ is $d$. Then
  there are exactly $\varphi(d_j)$ elements of order $d_j$ modulo
  $p_j^{e_j}$.
\end{proof}

Let us have a look of this behavior on an example:
let $n=45$ so that $\varphi(45) = 6 \times 4 = 24$ and 
$\lambda(45) = 12$. We thus know that any order modulo $9$ divides
$\varphi(9) = 6$ and that any order modulo $5$ divides 
$\varphi(5) = 4$. This gives the different orders of the $24$
invertibles shown on table \ref{tab:45}.
\begin{table}[ht]
\begin{center}\begin{small}\begin{tabular}{|lccr|}
\hline
\multicolumn{3}{|c}{order} & \# of elements of that\\
modulo 45  &    modulo 9       &  modulo 5        & order modulo 45\\
\hline
1     & 1 & 1 & \bf{1}\\
\hline
  &  1 & 2 & 1\\
  &  2 & 1 & 1\\
  &  2 & 2 & 1\\
  & \multicolumn{3}{r|}{\rule{60mm}{0.1pt}}\\
2 &    &   & \bf{3}\\
\hline
3
 & 3 & 1 & $\varphi(3)  \times \varphi(1) = $ \bf{2}\\
\hline
 & 1 & 4 & $ \varphi(1) \times \varphi(4) = 2$\\
  & 2 & 4 & $\varphi(2) \times \varphi(4) = 2$\\ 
  & \multicolumn{3}{r|}{\rule{60mm}{0.1pt}}\\
4 &   &   & \bf{4}\\
\hline
  &  6 & 1 & $\varphi(6) \times \varphi(1) = 2$\\
  &  3 & 2 & $\varphi(3) \times \varphi(2) = 2$\\
  &  6 & 2 & $\varphi(6) \times \varphi(2) = 2$\\
  & \multicolumn{3}{r|}{\rule{60mm}{0.1pt}}\\
6 &    &   & \bf{6}\\
\hline
  & 3 & 4 & $\varphi(3) \times \varphi(4) = 4$\\
  & 6 & 4 & $\varphi(6) \times \varphi(4) = 4$\\ 
  & \multicolumn{3}{r|}{\rule{60mm}{0.1pt}}\\
12 &   &   & \bf{8}\\
\hline
\end{tabular}\end{small}
\caption{Elements of a given order modulo 45}\label{tab:45}
\end{center}
\end{table}%\vspace{-5ex}
It would be highly desirable to have tight bounds on those number of
elements of a given order. Moreover, these bounds should be easily computable
(e.g. not requiring some factorization !). 
In \cite{Cameron:2003:lambdaroots,Muller:2004:lambdaroots}, the following is proposed:
\begin{proposition}\label{cj:lambda}\cite[Corollary 6.8]{Cameron:2003:lambdaroots}
For $n$ odd, the number of elements of order $\lambda(n)$ (primitive
$\lambda-$roots) is larger than 
$\varphi(\varphi(n))$.
\end{proposition}
%The bound is tight: $\varphi(\varphi(15))=4$
%and only $2$, $7$, $8$ and $13$ have order $\lambda(15)=4$.
%On the other hand,
%for example, $\varphi(\varphi(21))=4$ but more than $4$ elements ($2$,
%$5$, $10$, $11$, $17$ and $19$) are of order $\lambda(21)=6$. 
%
Now, this last result shows that actually quite a lot of elements are of
maximal order modulo $n$. Using this fact, a modification of algorithm
$1$ can then produce with high probability an element of maximal order
even though $n$ is composite.
%
%
%%%%%%%%%%%%%%%%%%%%%%%%%%%%%%%%%%%%%%%%%%%%%%%%%%%%%%%%%%%%
%%% Local Variables:
%%% mode: latex
%%% TeX-master: "polypr"
%%% End:
 %% tentative analysis for composite numbers
%
%\newpage
\section{Applications}\label{sec:applis}
Of course, our generation can be applied to any application requiring
the use of primitive roots. In this section we show the speed of our
method compared to generation of primes with known factorization and
propose a generalization of Miller-Rabin probabilistic primality test
and
of Davenport's strengthenings \cite{Davenport:1992:PTR}.
%\newpage
\subsection{Faster pseudo random generators construction or key
  exchange}
\begin{figure}[htb]
\includegraphics[width=\columnwidth]{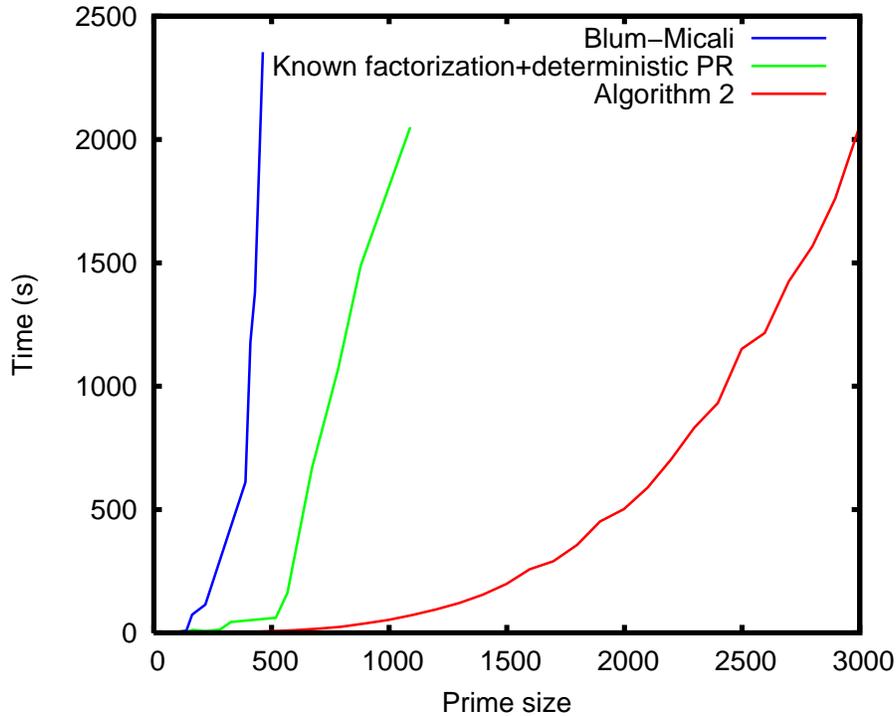}
\caption{Blum-Micali primes with known
factorization vs Industrial-strength primitive roots}\label{fig:pwnf}
\end{figure}

The use of a generator and a big prime is the core of many
cryptographic protocols. Among them are Blum-Micali pseudo-random
generators \cite{Blum:1984:BMSSPR}, Diffie-Hellman key exchange
\cite{Diffie:1976:NDC}, etc.\\
In this section we just compare the generation of primes with known
factorization \cite{Bach:1988:HGF}, 
so that primitive roots of primes with any given size
are computable. 
The idea in \cite{Blum:1984:BMSSPR} is to iteratively and randomly 
build primes so that
the factorizations of $p_i - 1$ are known.
For cryptanalysis reasons their original method selects the primes
and primitive roots bit by bit and is therefore quite slow. 
On figure
\ref{fig:pwnf} we then present also a third way, which is to generate
the prime with known factorization as in \cite{Bach:1988:HGF}, but
then to generate the primitive root deterministically with our
algorithm (since the factorization of $p-1$ is known).
We compare this method with the following full-probabilistic way: 
\begin{enumerate}\vspace{-2ex}
\item By trial and error generate a probable prime (e.g. a prime
  passing several Miller-Rabin tests \cite{Miller:1975:Riemann}).
\item Generate a probable primitive root by Heuristic 2.
%Algorithm \ref{algobest}.
\end{enumerate}
We see on figure \ref{fig:pwnf} that our method is faster and allows
for the use of bigger primes/generators. 
%
%\newpage
\subsection{Probabilistic Lucas primality test}\label{ssec:lucas}
The deterministic primality test of Lucas is actually the existence of
primitive roots:
\begin{theorem}[Lucas]\label{thm:Lucas}
Let $p>0$. If one can find an $a>0$ such that 
 $a^{p-1}\equiv 1\,mod\,p$ and
$a^{\frac{p-1}{q}}\nequiv 1\,mod\,p$, as soon as $q$ divides $p-1$, then 
$p$ is prime.
\end{theorem}
We propose here as a probabilistic primality test to try to build a
primitive root. If one succeeds then the number is prime with high
probability else it is either proven composite or composite with a
high probability.\\
Now for the complexity, we do not pretend to challenge Miller-Rabin
test for speed ! Well, one often needs to perform several
Miller-Rabin tests with distinct witnesses, so that the probability of
being prime increases. Our idea is the following: since one tests
several witnesses, why not use them as factors of our probable
primitive root ! 
This idea can then be viewed as a generalization of
Miller-Rabin: we not only test for orders of the form
$\frac{n-1}{2^e}$ but also for each order of the form
$\frac{n-1}{q^e}$ where $q$ is a small prime factor of $n-1$.
The effective complexity (save maybe from the partial
factorization) will not suffer and the probability can jump as
soon as an element with very high order is generated.
The algorithm is then a slight modification of algorithm \ref{algo1},
where we let $F(B,Q)=1-(1+\frac{1}{Q-1})(1-\frac{1}{B})^{\log_B{Q}}$:
\begin{algorithm2e}[htb]\caption{Probabilistic Lucas primality test}
\raggedright
\label{algo:lucas}
  \KwIn{$n\geq 3$, odd.}
  \KwIn{A failure probability $0 < \epsilon < 1$.}
  \KwOut{\KwWhether $n$ is prime and a certificate of primality,}
  \KwOut{\KwOr $n$ is composite and a factor (or just a Fermat witness),}
  \KwOut{\KwOr $n$ is prime with probability of error less than
    $\epsilon$,}
  \KwOut{\KwOr $n$ is composite with probability of error less than
    $\epsilon$.}
  \Begin{
    Set $P = 1$, $a = 1$, $Q=n-1$ and $q=2$.\\
    \While{$Q>n^{\frac{2}{3}}$}{
      Randomly choose $\alpha \mod n$.\\
      \gIf{$gcd(\alpha,n) \neq 1$
        \KwOr 
      $gcd(\alpha^{\frac{n-1}{q}}-1,n)  \notin \{1;n\}$ 
        \KwOr
      $\alpha^{n-1} \nequiv 1 [n]$
        \KwOr  
      $( q==2$
                \KwAnd
      $n~\text{is not a
          strong pseudoprime to the base}~\alpha)$
      }{ 
        \KwRet{$n$ is composite.}
      }\gElsIf{$\alpha^{\frac{n-1}{q}} \equiv 1\;mod\;n$}{
        Set $P = P / q$.\\
        \If{$P \leq \epsilon$}{
          \KwRet{$n$ is probably composite with error less than $P$.}
        }
      }\gElse{
        {\bf -} Set $e$ to the greatest power of $q$ dividing $Q$.\\
        {\bf -} Set $Q = Q / q^e$.\\
        {\bf -} Set $a = a \times \alpha^{\frac{n-1}{q^e}}$.\\
        {\bf -} Set $k = k \cup \lbrace q^e \rbrace$.\\
        {\bf -} Refine $B$ such that $F(B,Q) == 4\epsilon$.\\
        {\bf -} Find a new prime factor $q$ of $Q$ with $q<B$,
        otherwise set $q = Q$.
      }
    }
    \eIf{Every $q$ was prime}{
      \KwRet{$n$ is prime and $(a,k)$ is a certificate.} 
    }{
      \KwRet{$n$ is probably prime with error less than $F(B,q)$.}
    }
  }
\end{algorithm2e}
\begin{remark}The exponentiations by $\frac{n-1}{q}$ can in practice
  be factorized in a ``Lucas-tree''
  \cite{Pratt:1975:EPSC,Crandall:2001:PNaCP}.
\end{remark}
\begin{remark} Algorithm \ref{algo:lucas} is correct for the primes
  and most of the composite numbers.
\end{remark}
\begin{proof} Correctness for prime numbers is the correctness of the
  pseudo primitive root generation. \\
Now for composite numbers: the idea is that first of all, only
Carmichael numbers will be able to pass the pseudo prime test several
times.\\
 The $4\epsilon$ then follows since at least one
$\alpha$ passed the strong pseudoprime test. This reduces the possible
Carmichael numbers able to pass our test.
Then, for most of the Carmichael numbers, 
$\lambda(n)$ divides $n-1$ but, moreover, $\lambda(n)$ also divides
$\frac{n-1}{q}$ for some $q$, factor of $n-1$. Therefore,
$\alpha^{\frac{n-1}{q}}$ will always be one. If $n$ is prime on the
contrary, only $\frac{1}{q}$ elements will have order a multiple of
$q$.\\
Now for the $n^{\frac{2}{3}}$ in the loop. The argument is the same
as for the Pocklington  theorem 
\cite[Theorem 4.1.4]{Crandall:2001:PNaCP} and the 
Brillhart, Lehmer and Selfridge theorem
\cite[Theorem 4.1.5]{Crandall:2001:PNaCP}: let $n-1=kQ$ and let $p$ be
a prime factor of $n$. The algorithm has found an $a$ verifying
$a^{n-1} \equiv 1 \mod n$. Hence, the order of $a^Q \mod p$ is a
divisor of $\frac{n-1}{Q} = k$. Now, since 
$gcd(a^{\frac{n-1}{q}}-1,n) = 1$ for each prime $q$ dividing $k$, this
order is not a proper divisor of $k$, so is equal to $k$. Hence, $k$
must be a divisor of $p-1=\varphi(p)$. We conclude that each prime
factor of $n$ must exceed $k$. From this, Pocklington's theorem states that
if $k$ is greater than $\sqrt{n}$, $n$ is prime. And then,
Brillhart-Lehmer-Selfridge theorem states that if $k$ is in between
$n^{\frac{1}{3}}$ and $n^{\frac{1}{2}}$ then $n$ must be prime or
composite with exactly two prime factors 
\cite[Theorem 4.1.5]{Crandall:2001:PNaCP}. But $n$ has escaped our
previous tests only if $n$ is a Carmichael number. Fortunately,
Carmichael numbers must have at least $3$ factors 
\cite[Proposition V.1.3]{Koblitz:1987:CNT}. Now, whenever $Q$ is
below $n^{\frac{2}{3}}$, $k$ exceeds $n^{\frac{1}{3}}$ and then $n$
must be prime otherwise $n$ would have more than $3$ factors each of
those being greater than $n^{\frac{1}{3}}$.
\end{proof}

\begin{figure}[htb]
\includegraphics[width=\columnwidth]{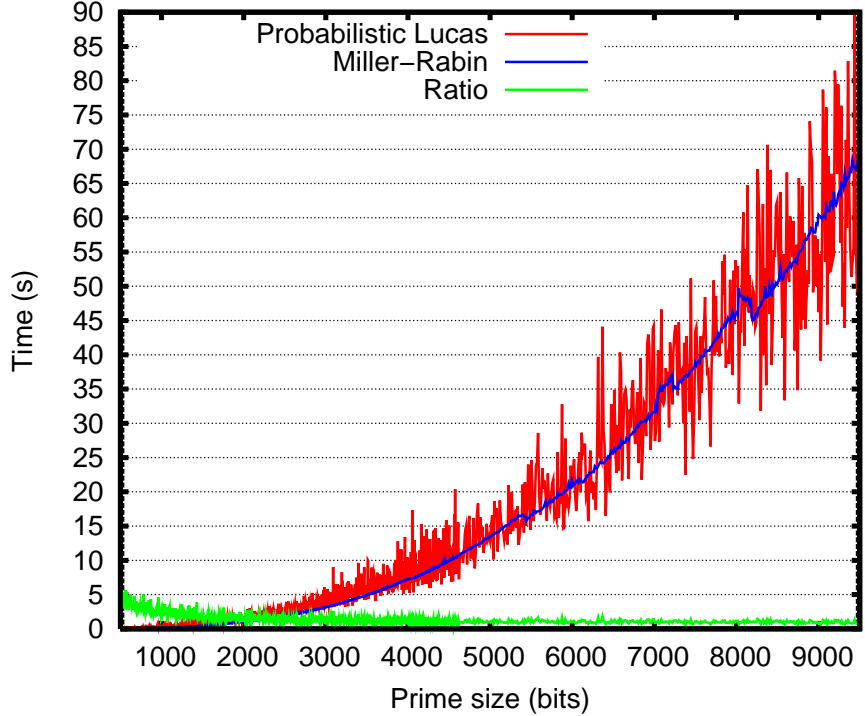}
\caption{Probabilistic Lucas vs GMP's Miller-Rabin for primes with probability $< 10^{-6}$}\label{fig:plp}
\end{figure}

Here is an example of Carmichael number, $1729$. $1728 = 2^6 3^3$,
where $\lambda(1729) = 2^2 3^2$. Then $\frac{n-1}{q}$ is either $864$
or $576$ both of which are divisible by 
$36 = \lambda(1729)$. 
Therefore, our test will detect $1729$ to be probably composite
with any probability of correctness.
Figure \ref{fig:plp} shows that this algorithm is highly competitive
with repeated applications of GMP's strong pseudo prime test
(i.e. with the same estimated probability of correctness).
Depending on the success of the partial factorization, our test can even
be faster (timing, on a PIV 2.4GHz, presented on figure \ref{fig:plp} are the mean time
between 4 distinct runs).\\
\nocite{Friedlander:2002:CPP}
Haplessly, some Carmichael numbers will still pass our test. The
following results, sharpening \cite[lemma 1]{Friedlander:2001:PPG}, 
explains why:
\begin{theorem}\label{theo:orddiv} Let $n=p_1^{e_1}\ldots p_\omega^{e_\omega}$. 
Let $q$ be a prime divisor of $\varphi(n)$, and
$(f_1,\ldots,f_\omega)$ be the maximal values for which
$q^{f_i}$ divides $\varphi(p_i^{e_i})$. 
There are $$\varphi(n) \left( 1 - \frac{1}{q^{\sum f_i}} \right)$$ invertible
  elements of order divisible by $q$ (i.e. for which
  $\alpha^{\frac{\lambda(n)}{q}} \nequiv 1 \mod n$).
\end{theorem}
\begin{proof}
By the Chinese remainder theorem, one can consider the moduli by
$p_i^{e_i}$ separately. 
Suppose, without loss of generality, that
$p_1^{e_1}$ is such that $f_1 > 0$. Otherwise all the $f_i$ are $0$
and the theorem is still correct.
Consider a generator $g$ of
the invertibles modulo $p_1^{e_1}$. 
An element has $q$ in its
order if and only if its index with respect to $g$ contains $q^{f_1}$.
There are exactly $1-\frac{1}{q^{f_1}}$ such elements among the
elements of $\pF{p_1^{e_1}}$. By the Chinese theorem, 
among the elements having their order
divisible by $q$ modulo $n$, we have then identified 
$\varphi(n) ( 1-\frac{1}{q^{f_1}})$ of them: the ones having
their order modulo $p_1^{e_1}$ divisible by $q$. 
Now the others are among the 
$\varphi(n) ( \frac{1}{q^{f_1}})$ that remains. Just now
consider those modulo $p_2^{e_2}$. If $f_2 == 0$ then we have not
found any new element. Otherwise, $1-\frac{1}{q^{f_2}}$ of them are of
order divisible by $q$. Well, actually, 
in both cases, we can state that $1-\frac{1}{q^{f_2}}$ of them are of
order divisible by $q$. We have thus found some other elements:
$\varphi(n) (
  \frac{1}{q^{f_1}})(1-\frac{1}{q^{f_2}})$. This
added to the previously found elements makes 
$\varphi(n) ( 1-\frac{1}{q^{f_1}q^{f_2}})$. Doing such a
counting for each of the remaining $p_i^{e_i}$ gives the announced
formula.
\end{proof}

For instance, take a Carmichael number still passing our test whenever 
$B~\leq~1450$: $37690903213=229 \times 2243 \times 73379$.
 Well, 
$37690903212 = 19 \times 2^2 \times 3 \times 59 \times 1451 \times
1931$ and 
$\lambda(37690903213)=19 \times 2^2 \times 3 \times 59 \times 1931$.
Then, $Q$ will be $1451 \times 1931$ and our algorithm will be able to
find elements for which $\alpha^{\frac{n-1}{Q}} \nequiv 1 \mod n$: 
those of which order is divisible by $1931$. Unfortunately, there are
quite a lot of them: $\varphi(n)\frac{1930}{1931} = 37489647840
\approx (1-.00533962722683134975)n$. Thus, there are more than 5
chances over a thousand to choose an element $\alpha$ for which
$\alpha^{\frac{n-1}{1451 \times 1931}} \nequiv 1 \mod n$. 
Even though this is much higher than $\frac{1}{Q}$ (if $n$ was
prime), this probability will not be detected abnormal by our algorithm.
Now, even if $p-1$ is seldom smooth for $p$ prime \cite{Pomerance:2002:SOC},
one can wonder if this is still the case for this special kind of 
Carmichael numbers $\ldots$
%
%But  there is only $12282952320$ elements of order 
%$\lambda(37690903213)  = 25975812$, which is only $\approx =0.3258864$.
%
%%%%%%%%%%%%%%%%%%%%%%%%%%%%%%%%%%%%%%%%%%%%%%%%%%%%%%%%%%%%
%%% Local Variables:
%%% mode: latex
%%% TeX-master: "polypr"
%%% End:
 %% Generators, Primality testing
%
%\newpage
\section{Conclusion}
We provide here a new very fast and efficient algorithm 
generating primitive roots.
On the one hand, the algorithm has a polynomial time bit complexity
when all existing
algorithms where exponential. This is for instance illustrated when
comparing it to existing software on figure \ref{fig:pari}.
On the other hand, our algorithm is probabilistic in the sense that the
answer might not be a primitive root. We have seen in this paper
however, that
the chances that an incorrect answer is given are less important than
say ``hardware malfunctions''. For this reason, we call our answers
``Industrial-strength'' primitive roots.\\
% An application of our algorithm is a new primality
% test: {\em attempt to construct a primitive root
% for a number $n$ with algorithm \ref{algo1} ;
% if $n$ is prime, the algorithm will succeed, 
% otherwise the $\alpha_i^{\frac{p-1}{p_i}}$ will be $1$
% too often \cite{jgd:2004:RTpr}}.
% When a given probability of success is required, this
% algorithm can be competitive with repeated applications of
% Miller-Rabin's test and allows to quantify the
% information gained by finding elements of large order.

 Then, we propose a new probabilistic primality test using this
 primitive root generation. This test can be viewed as a generalization of
 Miller-Rabin's test to other small prime factors dividing $n-1$
 The test is then quantifying the
 information gained by finding elements of large order modulo $n$.
 When a given probability of correctness is desirable for the test, our
 algorithm is heuristically competitive with repeated applications of
 Miller-Rabin's.
\section*{Acknowledgements}
Many thanks to T. Itoh and E. Bach.
%
%
% {\footnotesize
\bibliographystyle{plain}
\bibliography{jgdbibl}

\begin{thebibliography}{10}

\bibitem{Bach:1988:HGF}
Eric Bach.
\newblock How to generate factored random numbers.
\newblock {\em SIAM Journal on Computing}, 17(2):179--193, April 1988.
\newblock Special issue on cryptography.

\bibitem{Bach:1997:sppr}
Eric Bach.
\newblock Comments on search procedures for primitive roots.
\newblock {\em Mathematics of Computation}, 66(220):1719--1727, October 1997.

\bibitem{Bach:1996:ANTEA}
Eric Bach and Jeffrey Shallit.
\newblock {\em Algorithmic Number Theory: Efficient Algorithms}.
\newblock MIT press, 1996.

\bibitem{Blum:1984:BMSSPR}
Manuel Blum and Silvio Micali.
\newblock How to generate cryptographically strong sequences of pseudo-random
  bits.
\newblock {\em {SIAM Journal on Computing}}, 13(4):850--864, November 1984.

\bibitem{Cameron:2003:lambdaroots}
Peter~J. Cameron and D.~A. Preece.
\newblock Notes on primitive $\lambda-$roots, March 2003.
\newblock \url{http://www.maths.qmul.ac.uk/~pjc/csgnotes/}\url{lambda.pdf}.

\bibitem{Crandall:2001:PNaCP}
Richard Crandall and Carl Pomerance.
\newblock {\em Prime Numbers, a computational perspective}.
\newblock Springer, 2001.

\bibitem{Davenport:1992:PTR}
J.~H. Davenport.
\newblock Primality testing revisited.
\newblock In Paul~S. Wang, editor, {\em Proceedings of {ISSAC} '92.
  International Symposium on Symbolic and Algebraic Computation}, pages
  123--129, New York, NY 10036, USA, 1992. ACM Press.

\bibitem{Diffie:1976:NDC}
Whitfield Diffie and Martin~E. Hellman.
\newblock New directions in cryptography.
\newblock {\em IEEE Transactions on Information Theory}, IT-22(6):644--654,
  1976.

\bibitem{Elliott:1997:ALPR}
Peter D. T.~A. Elliott and Leo Murata.
\newblock On the average of the least primitive root modulo p.
\newblock {\em Journal of The london Mathematical Society}, 56(2):435--454,
  1997.

\bibitem{Erdos:1991:CLF}
Paul Erd{\"o}s, Carl Pomerance, and Eric Schmutz.
\newblock Carmichael's lambda function.
\newblock {\em Acta Arithmetica}, 58:363--385, 1991.

\bibitem{Friedlander:2001:PPG}
John~B. Friedlander, Carl Pomerance, and Igor Shparlinski.
\newblock Period of the power generator and small values of {Carmichael}'s
  function.
\newblock {\em Mathematics of Computation}, 70(236):1591--1605, October 2001.
\newblock See corrigendum \cite{Friedlander:2002:CPP}.

\bibitem{Friedlander:2002:CPP}
John~B. Friedlander, Carl Pomerance, and Igor Shparlinski.
\newblock Corrigendum to {``Period of the power generator and small values of
  Carmichael's function''}.
\newblock {\em Mathematics of Computation}, 71(240):1803--1806, October 2002.
\newblock See \cite{Friedlander:2001:PPG}.

\bibitem{VonzurGathen:1998:OGP}
Joachim~{von zur} Gathen and Igor Shparlinski.
\newblock Orders of {Gauss} periods in finite fields.
\newblock {\em Applicable Algebra in Engineering, Communication and Computing},
  9:15--24, 1998.

\bibitem{Hardy:1979:ITN}
Godfrey~Harold Hardy and E.~Maitland Wright.
\newblock {\em An Introduction to the Theory of Numbers}.
\newblock Oxford University Press, fifth edition, 1979.

\bibitem{Itoh:2001:PPR}
Toshiya Itoh and Shigeo Tsujii.
\newblock How to generate a primitive root modulo a prime.
\newblock Technical Report 009-002, IPSJ SIGNotes ALgorithms Abstract, 2001.

\bibitem{Knuth:1997:TAoCPSA}
Donald~E. Knuth.
\newblock {\em Seminumerical Algorithms}, volume~2 of {\em The Art of Computer
  Programming}.
\newblock Ad{\-d}i{\-s}on-Wes{\-l}ey, Reading, MA, USA, $2^{nd}$ edition, 1997.

\bibitem{Koblitz:1987:CNT}
Neal Koblitz.
\newblock {\em A course in number theory and cryptography}, volume 114 of {\em
  Graduate texts in mathematics}.
\newblock Spring{\-}er-Ver{\-}lag, Berlin, Germany~/ Heidelberg, Germany~/
  London, UK~/, etc., 1987.

\bibitem{Miller:1975:Riemann}
Gary~L. Miller.
\newblock Riemann's hypothesis and tests for primality.
\newblock In {\em Conference Record of Seventh Annual {ACM} Symposium on Theory
  of Computation}, pages 234--239, Albuquerque, New Mexico, May 1975.

\bibitem{Muller:2004:lambdaroots}
Thomas~W. M{\"u}ller and Jan-Christof Schlage-Puchta.
\newblock On the number of primitive $\lambda-$roots.
\newblock {\em Acta Arithmetica}, 115(3):217--223, 2004.

\bibitem{OGorman:1996:FTC}
T.~J. O'Gorman, J.~M. Ross, A.~H. Taber, J.~F. Ziegler, H.~P. Muhlfeld, C.~J.
  Montrose, H.~W. Curtis, and J.~L. Walsh.
\newblock Field testing for cosmic ray soft errors in semiconductor memories.
\newblock {\em IBM Journal of Research and Development}, 40(1):41--50, January
  1996.

\bibitem{Pomerance:2002:SOC}
Carl Pomerance and Igor~E. Shparlinski.
\newblock Smooth orders and cryptographic applications.
\newblock {\em Lecture Notes in Computer Science}, 2369:338--348, 2002.
\newblock ANTS-V: 5th International Algorithmic Number Theory Symposium.

\bibitem{Pratt:1975:EPSC}
Vaughan~R. Pratt.
\newblock Every prime has a succinct certificate.
\newblock {\em SIAM Journal on Computing}, 4(3):214--220, 1975.

\bibitem{Robin:1983:omega}
Guy Robin.
\newblock Estimation de la fonction de {Tchebycheff} $\theta$ sur le k-i\`eme
  nombre premier et grandes valeurs de la fonction $\omega(n)$ nombre de
  diviseurs premiers de $n$.
\newblock {\em Acta Arithmetica}, XLII:367--389, 1983.

\bibitem{Shoup:1992:SPR}
Victor Shoup.
\newblock Searching for primitive roots in finite fields.
\newblock {\em Mathematics of Computation}, 58(197):369--380, January 1992.

\bibitem{Wagstaff:2003:CNTC}
Samuel~S. Wagstaff, Jr.
\newblock {\em Cryptanalysis of number theoretic ciphers}.
\newblock Chapman-Hall / CRC, 2003.

\end{thebibliography}
% }
% {\footnotesize
% \input{jgdppr.bbl}
% }

\end{document}